\pdfoutput=1
\documentclass[twocolumn]{aastex63}

\shorttitle{Orion Expansion}
\shortauthors{Swiggum et al.}

\usepackage{float}
\usepackage{amsmath,bm}
\usepackage[utf8]{inputenc}
\usepackage[normalem]{ulem}

\begin{document}

\title{Evidence for Radial Expansion at the Core of the Orion Complex with \textit{Gaia} EDR3}

\correspondingauthor{Cameren Swiggum}
\email{swiggum2@wisc.edu}

\author[0000-0001-9201-5995]{Cameren Swiggum}\affiliation{Department of Astronomy, University of Wisconsin, 475 North Charter Street, Madison, WI 53706, USA }

\author{Elena D'Onghia}\affiliation{Department of Astronomy, University of Wisconsin, 475 North Charter Street, Madison, WI 53706, USA }

\author[0000-0002-4355-0921]{João Alves}\affiliation{University of Vienna, Department of Astrophysics, Türkenschanzstraße 17, 1180 Vienna, Austria}

\author[0000-0002-0568-5526]{Josefa Gro{\upshape{\ss}}schedl}\affiliation{University of Vienna, Department of Astrophysics, Türkenschanzstraße 17, 1180 Vienna, Austria}

\author[0000-0002-6747-2745]{Michael Foley}\affiliation{Center for Astrophysics $\vert$ Harvard $\&$ Smithsonian, 60 Garden St., Cambridge, MA 02138, USA}

\author[0000-0002-2250-730X]{Catherine Zucker}\affiliation{Center for Astrophysics $\vert$ Harvard $\&$ Smithsonian, 60 Garden St., Cambridge, MA 02138, USA}

\author[0000-0002-0568-5526]{Stefan Meingast}\affiliation{University of Vienna, Department of Astrophysics, Türkenschanzstraße 17, 1180 Vienna, Austria}

\author{Boquan Chen}\affiliation{Sydney Institute for Astronomy, The University of Sydney, School of Physics A28, Camperdown, NSW 2006, Australia }

\author[0000-0003-1312-0477]{Alyssa Goodman}\affiliation{Center for Astrophysics $\vert$ Harvard $\&$ Smithsonian, 60 Garden St., Cambridge, MA 02138, USA}

\begin{abstract}
We present a phase-space study of two stellar groups located at the core of the Orion complex: Brice{\~n}o-1 and Orion Belt Population-near (OBP-near). We identify the groups with the unsupervised clustering algorithm, Shared Nearest Neighbor (SNN), which previously identified twelve new stellar substructures in the Orion complex. For each of the two groups, we derive the 3D space motions of individual stars using \textit{Gaia} EDR3 proper motions supplemented by radial velocities from \textit{Gaia} DR2, APOGEE-2, and GALAH DR3. We present evidence for radial expansion of the two groups from a common center. Unlike previous work, our study suggests that evidence of stellar group expansion is confined only to OBP-near and Briceño-1, whereas the rest of the groups in the complex show more complicated motions. Interestingly, the stars in the two groups lie at the center of a dust shell, as revealed via an extant 3D dust map. The exact mechanism that produces such coherent motions remains unclear, while the observed radial expansion and dust shell suggest that massive stellar feedback could have influenced the star formation history of these groups. 
\end{abstract}


\section{Introduction}
The formation and evolution of star clusters is a complex process that depends crucially on the formation of individual stars and how feedback mechanisms, including winds, radiation pressure, and supernovae, affect their environment. Young stars are usually grouped in clusters and are located within their natal star-forming regions. In contrast to younger stars, older stars are found dispersed throughout the Galactic field. 

The role of gas in the formation and evolution of bound clusters or associations of stars is still poorly understood (e.g., \citealt{krause2020}). A popular scenario posits that stars form and temporarily persist in dense molecular clouds, held together by the gravitational potential of the remaining gas \citep{lada2003}. The gas is eventually blown-out by feedback from stellar winds and supernovae events. Stars may disperse due to this gas expulsion and the associated change in gravitational potential \citep{tutukov1997, hills1980, goodwin2006, baumgardt2007, krause2020}. Recent simulations also suggest that the expelled gas can flip a stellar association's gravitational potential and elicit ``gravitational feedback'', accelerating the expansion \citep[e.g.,][]{krause2020, zamora2019}. Another model theorizes that once the gas is expelled, associations of stars might form by compression in the expanding shell \citep{elmegreen1977}.  

The Orion complex is the most massive star-forming region in the solar vicinity \citep{bally2008}.  Given its proximity \citep[$\approx$ 400 pc,][]{Zucker_2020,Grossschedl_2018}, its mass \citep[$>10^{5}$\text{ M$_{\odot}$;}][]{lada2010}, and the presence of multi-phase gas and stars at various evolutionary stages, Orion represents one of the best places to observe young stars in their natal environments. 

Before the \textit{Gaia} era, the classification of reliable associations and young clusters within the Orion complex was more challenging. While overdensities of stars were identified \citep[e.g.,][]{blaauw1964, brown1994, bally2008}, often these overdensities appeared superimposed on the plane of the sky. The lack of accurate distances and proper motions prevented a separation of distinct groups in true 3D space, making it difficult to classify the stellar associations in the Orion complex.

This situation has changed with the advent of the European Space Agency's (ESA) \textit{Gaia} Mission, which has provided distances and proper motions for tens of thousands of young stars in the vicinity of Orion \citep{gaia2016}. These data, supplemented by radial velocities (RVs) from \textit{Gaia}, the Sloan Digital Sky Survey IV \citep[SDSS-IV;][]{majewski2017} APOGEE-2 \citep{blanton2017}, and GALAH DR3 \citep{bunder2020}, 
have allowed the use of clustering algorithms to identify stellar associations in phase-space, with implications for understanding how young clusters form and evolve.

Previous studies of the young, star-forming Orion Nebular Cluster \citep[ONC;][]{hillenbrand1998, jerabkova2019} reveal the impact of stellar feedback within the region \citep{kroupa2001,kroupa2018}. However, at a young age \citep[$\sim 2.5$ Myr;][]{jeffries_2011}, the bulk expansion of the region's stars is weak \citep{kuhn2019}. Other works explored the entire Orion complex and identified numerous stellar groups of various ages \citep[e.g.,][]{kubiak2017, kounkel2018, kos2019, zari2019}. A 6D phase-space analysis of the $\lambda$ Orion region has shown evidence for expansion \citep{kounkel2018}. Additionally, a gas study of Orion has also shown evidence for expansion likely due to strong stellar feedback \citep{josefa2020}.

\cite{chen2019} applied an unsupervised clustering algorithm to Gaia DR2 data and identified 21 stellar groups in the Orion Complex. While nine of them overlap with stellar groups previously identified in the Orion region \citep{briceno2007, alves2012, kounkel2018}, twelve were newly discovered. Recently, \cite{kounkel2018} cataloged the Orion region groups using a different clustering algorithm. \cite{kounkel2020} further analyzed the groups' phase space and claimed that the entire complex is expanding from a central region. Here we show that the expansion is most evident in two massive stellar groups in Orion, namely OBP-near and Brice{\~n}o-1, which exhibit a ``Hubble flow'' like expansion pattern similar to those found in other star-forming regions \citep[e.g.,][]{wright2018,wright2019,kuhn2019,kuhn2020, cantat2019,roman2019}. Our analysis uses \textit{Gaia} EDR3, APOGEE-2, and GALAH DR3 to characterize the stars in the phase space and determine the age of the two groups of which OBP-near is newly discovered in a previous work \citep{chen2019}. 

In Section \ref{sec:data}, we layout the data and methodology used to identify the stellar groups and study their dynamics and ages. Section \ref{sec:results} reports the main results of our study. Section \ref{sec:discussion} discusses physical scenarios that could lead to our results and compares with previous works. Our results are briefly summarized in Section \ref{sec:conclusion}.

\section{Data Analysis}
\label{sec:data}
    \subsection{SNN Stellar Groups}
    
        \cite{chen2019} applied two clustering algorithms, namely Shared Nearest Neighbor \citep[SNN;][]{chen2018} and \textit{EnLink} \citep{sharma2009}, to {\it Gaia} DR2
        in the vicinity of Orion, to uncover stellar groups clustered in RA, DEC, parallaxes ($\alpha, \delta, \varpi$) and proper motions ($\mu_{\alpha}, \mu_{\delta}$).    
        Advantages of this approach and its applied analysis include: 1) the ability to omit stars that unlikely belong to any group (unclassified stars); 2) the use of a  density threshold criterion to balance the high and low-density regions in the 5D-space; 3) the assignment of group stability scores based on how many times the star appears in each of the total iterations \citep[7,000 in][]{chen2019} and group membership probability.
        
        \begin{figure*}[!ht]
            \centering
            \makebox[\textwidth]{\includegraphics[width=
            0.9\paperwidth]{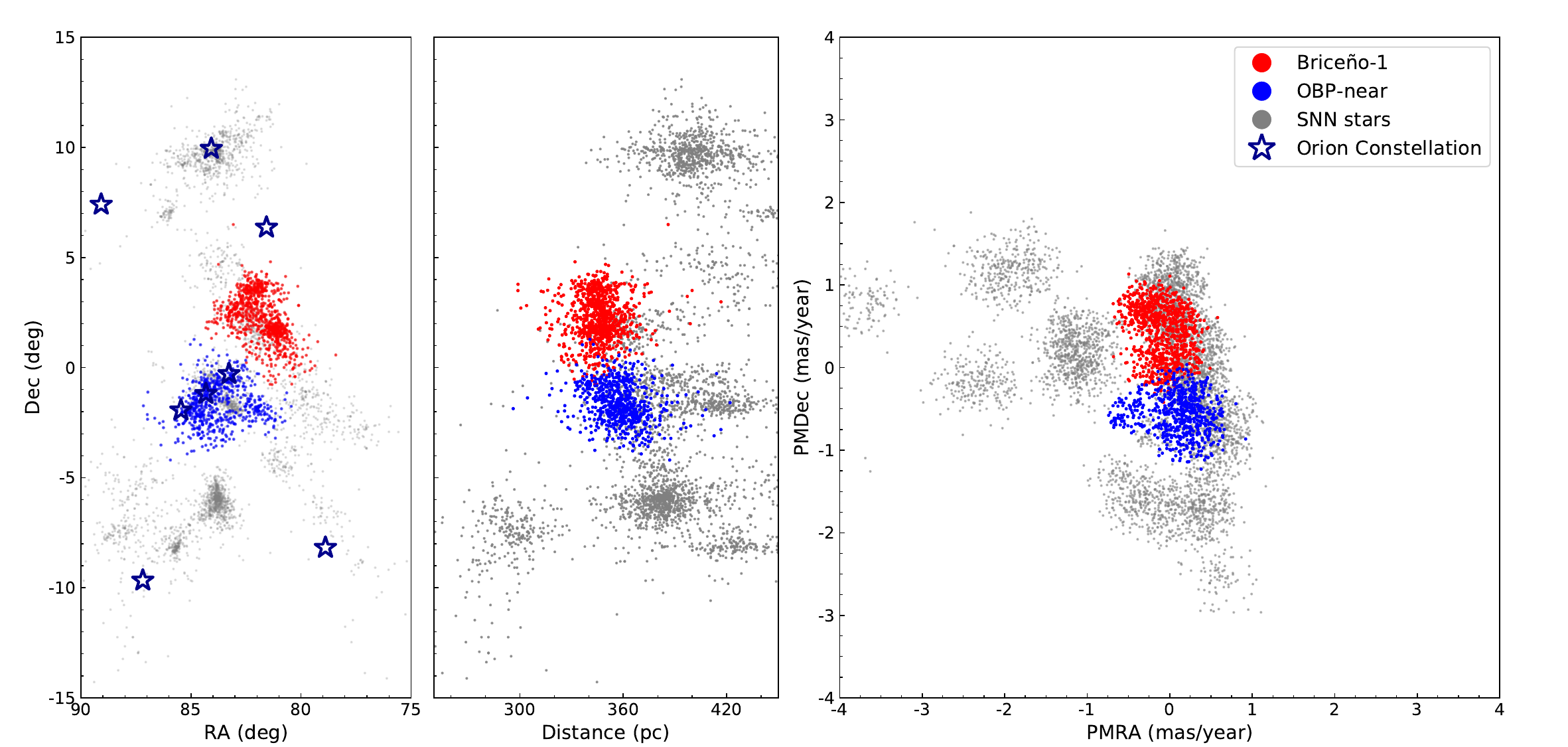}}
            \caption{Overview of stellar overdensities recovered by SNN in this work. Colored points show SNN members of OBP-near and Briceño-1 whereas gray points show SNN members of the other recovered groups. From left to right the panels display: (left) the group members' sky positions with star-shaped symbols marking the constellation of Orion; (center) distance versus declination; (right) locations in proper motion space.}
            \label{fig:orion_overview}
        \end{figure*}

        \indent Our study focuses on two groups labeled as SNN 1 and SNN 3 in \cite{chen2019}. SNN 1 is named ``Brice{\~n}o-1'' due to its significant overlap with the 25 Orionis stellar population studied by \cite{briceno2007}.\footnote{The prominent B star 25 Ori is not a member of SNN 1, hence the differing choice in the group's name.} The group is the most stable of all groups recovered in the original SNN analysis with a stability score of 3393 out of 7000, meaning it was recovered in $48.5 \%$ of the iterations (hereafter referred to as the stability percent). The plane-of-sky region of this stellar population is notably classified as the Ori OB1a sub-association and its average distance from the Sun is 350 pc. SNN 3 is named ``OBP-near'' and is newly discovered by the SNN clustering algorithm. It lies in the Ori OB1b sub-association \citep{blaauw1964} in the plane of the sky. OBP-near has a stability score of 2847 stability percent of $40.6 \%$, making it the third most stable group discovered in the SNN analysis. Its average distance from the Sun is 360 pc, establishing itself at the front of the Orion Belt Population, hence the distinguishing title ``near".
        
        Compared to \textit{Gaia} DR2, EDR3 shows $\sim 30 \%$ improvement in parallax precision and an improvement in proper motions by a factor of $\sim 2$ \citep{gaiaEDR3}. We query the \textit{Gaia} EDR3 database using the same region selection for Orion as used in \cite{chen2019}, specifically as $75^{\circ} < \alpha < 90^{\circ}$, $-15^{\circ}< \delta < 15^{\circ}$, $2 < \varpi < 5 \text{ mas}$, $-4< \mu_{\alpha} < 4 \text{ mas/yr}$, and $-4< \mu_{\delta} < 4 \text{ mas/yr}$, with a cut of $\varpi/\sigma_{\varpi} > 5$, which yields $33,811$ stars. We perform 500 iterations of SNN, each adopting a range of values for the free parameters  $n_{xyz}$ and $d_{pm}$. The $n_{xyz}$ parameter defines the number of nearest neighbors in 3D Cartesian position-space computed for each star. The $d_{pm}$ parameter is the maximum difference in proper motions between a star and its neighbors, computed for each star. We adopt uniform distributions of $n_{xyz}$ and $d_{pm}$ between 10--800 pc and 0--0.40 mas/yr, respectively. For the initial group finding, the parameters \textit{eps} and \textit{min\_samples} are fixed to the values \textit{eps}$=0.5$ and \textit{min\_samples}$=20$. This choice selects candidates with at least 20 stars having more than $50\%$ shared nearest neighbors. The step for finding reappearing groups across the 500 SNN runs uses \textit{eps}$=0.7$ and the minimum number of retrieved groups, \textit{min\_samples}, is set to 15 (see \cite{chen2019} for details). We adopt a probability cut of 2$\%$, meaning a star was included in a given group for at least 10 out of 500 iterations, greatly reducing the number of overlapped members. The average probability is 16$\%$ after this cut.

        Using EDR3, we retrieve Briceño-1 and OBP-near, along with thirteen other groups originally identified by \cite{chen2019} which have stability percents $\geq 5\%$ and include at least $50 \%$ of their original stars. Group names and basic properties are listed in Table \ref{tab:summary_table} and each recovered star is documented in Table \ref{tab:snn_catalog}. Briceño-1 and OBP-near astrometry are displayed in Figure \ref{fig:orion_overview} along with the other recovered SNN stars. Around $82\%$ of the original Briceño-1 stars are part of the new group with 349 new stars and $81\%$ of the original OBP-near stars are part of the new group with an additional 308 new stars. The subsets of stars recovered uniquely in this work or in \cite{chen2019} generally have lower SNN probabilities than their group neighbors, indicating their susceptibility to be classified as field stars in differing SNN runs. The newly recovered stars in this work are preferentially low mass stars with higher percent errors in astrometry when compared to their neighbors. As astrometric measurements continue to improve with future \textit{Gaia} data releases, recovery of low-brightness stars with SNN will become increasingly viable.
        
        Most importantly, we identify the original thirteen most stable groups \citep[stability percent $> 20 \%$;][]{chen2019} indicating that SNN is robust at identifying the most prominent overdensities in Orion. However, seven of the original low-stability groups \citep[stability $< 150;$][]{chen2019} are not recovered here. Additionally, five new, low-stability groups are recovered in our analysis which we do not document here. While the focus of this work pertains to the more massive, stable groups, future work will asses the more ambiguous stellar structure of Orion.

    \begin{deluxetable*}{cccccccccccccc}

\tablewidth{\columnwidth}
\tabletypesize{\scriptsize}
\tablecaption{Summary of Groups\label{tab:summary_table}}

\tablehead{\colhead{SNN Label} & \colhead{Name} & \colhead{Stability} & \colhead{N} & \colhead{Age} & \colhead{$A_V$} & \colhead{Percent DR2} & \colhead{$\bar{\alpha}$} & \colhead{$\bar{\delta}$} & \colhead{$\bar{d}$} & \colhead{$\bar{\mu_{\alpha}}$} & \colhead{$\bar{\mu_{\delta}}$} & \colhead{$\bar{\text{RV}}$} & \colhead{N$_{\text{RV}}$}
\\
& & & & (Myr) & (mag) & ($\%$) & (deg) & (deg) & (pc) & (mas~yr$^{-1}$) & (mas~yr$^{-1}$) & (km~s$^{-1}$) &
\\
(1) & (2) & (3) & (4) & (5) & (6) & (7) & (8) & (9) & (10) & (11) & (12) & (13) & (14)}

\rotate
\decimals
\startdata
\textbf{1} & \textbf{Briceño-1} & 383 & 909 & $9.0^{+4.0}_{-3.4}$ & 0.2 & 82.3 & $81.8~(0.9)$ & $2.2~(1.1)$ & $346.1~(11.5)$ & $1.4~(0.2)$ & $-0.2~(0.3)$ & $20.8~(10.6)$ & 147 \\
2 & Orion Y & 257 & 196 & $18.2^{+9.8}_{-5.3}$ & 0.4 & 84.3 & $87.9~(1.2)$ & $-7.6~(1.7)$ & $299.3~(21.7)$ & $0.1~(0.2)$ & $-0.6~(0.2)$ & $15.2~(11.4)$ & 21 \\
\textbf{3} & \textbf{OBP-near} & 192 & 656 & $6.8^{+3.6}_{-2.9}$ & 0.2 & 81.0 & $84.0~(1.2)$ & $-1.7~(0.9)$ & $358.3~(15.0)$ & $1.7~(0.3)$ & $-1.2~(0.2)$ & $22.9~(6.4)$ & 146 \\
4 & $\lambda$~Orion & 250 & 743 & $4.7^{+6.3}_{-2.4}$ & 0.4 & 79.4 & $83.8~(1.0)$ & $9.7~(0.8)$ & $400.4~(20.4)$ & $1.3~(0.4)$ & $-2.1~(0.2)$ & $27.7~(50.1)$ & 147 \\
5 & NGC 1980 & 137 & 773 & $4.1^{+6.1}_{-1.8}$ & 0.2 & 62.6 & $83.8~(0.4)$ & $-6.1~(0.6)$ & $383.2~(16.0)$ & $1.2~(0.2)$ & $0.5~(0.2)$ & $27.1~(5.4)$ & 197 \\
6 & OBP-d & 145 & 322 & $6.4^{+6.4}_{-4.3}$ & 0.3 & 79.3 & $83.1~(0.6)$ & $-1.7~(0.6)$ & $416.4~(20.8)$ & $0.1~(0.2)$ & $-0.2~(0.2)$ & $30.7~(8.1)$ & 41 \\
7 & ASCC20 & 85 & 200 & $22.5^{+5.0}_{-6.9}$ & 0.3 & 77.6 & $82.1~(0.7)$ & $1.7~(0.7)$ & $363.8~(18.5)$ & $-0.5~(0.3)$ & $0.8~(0.2)$ & $30.8~(76.8)$ & 36 \\
8 & L1616 & 158 & 272 & $5.9^{+4.3}_{-1.5}$ & 0.2 & 75.3 & $79.5~(1.2)$ & $-1.9~(1.0)$ & $366.9~(16.5)$ & $1.5~(0.2)$ & $-0.5~(0.2)$ & $22.2~(8.5)$ & 25 \\
9 & OBP-b & 96 & 229 & $17.9^{+6.1}_{-6.5}$ & 0.3 & 79.0 & $84.1~(0.7)$ & $-0.6~(0.6)$ & $385.2~(22.8)$ & $-1.1~(0.3)$ & $-0.7~(0.2)$ & $31.1~(5.8)$ & 29 \\
10 & $\lambda$~Orion South & 178 & 106 & $7.2^{+4.3}_{-4.4}$ & 0.6 & 74.4 & $86.0~(1.0)$ & $7.1~(1.1)$ & $440.6~(26.1)$ & $-3.3~(0.2)$ & $-1.7~(0.2)$ & $31.6~(17.7)$ & 3 \\
11 & Rigel & 157 & 85 & $10.8^{+3.8}_{-4.1}$ & 0.3 & 75.0 & $78.5~(1.5)$ & $-7.0~(2.2)$ & $285.3~(20.3)$ & $1.8~(0.2)$ & $-3.0~(0.2)$ & $14.6~(6.4)$ & 4 \\
12 & L1641S & 68 & 180 & $3.1^{+2.6}_{-2.1}$ & 0.3 & 58.0 & $85.7~(0.5)$ & $-8.1~(0.6)$ & $424.0~(23.6)$ & $0.1~(0.2)$ & $-0.2~(0.2)$ & $21.9~(5.1)$ & 46 \\
14 & ome Ori & 41 & 150 & $15.4^{+4.1}_{-4.2}$ & 0.6 & 67.0 & $83.5~(0.8)$ & $4.2~(1.0)$ & $411.8~(27.7)$ & $-0.9~(0.2)$ & $0.6~(0.2)$ & $32.1~(8.1)$ & 7 \\
21 & L1634 & 63 & 94 & $4.8^{+0.9}_{-2.4}$ & 0.4 & 83.1 & $81.0~(0.5)$ & $-4.3~(0.6)$ & $379.4~(17.0)$ & $1.9~(0.2)$ & $-1.0~(0.2)$ & $20.1~(3.2)$ & 2 \\
24 & Orion A East & 46 & 105 & $7.5^{+5.8}_{-4.0}$ & 0.6 & 71.9 & $87.5~(1.1)$ & $-5.2~(1.0)$ & $445.9~(30.2)$ & $-2.4~(0.2)$ & $0.3~(0.2)$ & $32.1~(13.5)$ & 5
\enddata

\tablecomments{Summary of SNN groups where each row reports the basic properties of an individual group. In column (1) we identify the original SNN group label assigned by \cite{chen2019}; (2) name of the group; (3) stability score from SNN; (4) number of stars in group; (5) age of group; (6) extinction of group; (7) percent of group's stars from \cite{chen2019} that are coincident with present work's stellar membership; (8) average/$1\sigma$ spread in RA; (9) average/$1\sigma$ spread in DEC; (10) average/$1\sigma$ spread in distance;  (11) average/$1\sigma$ spread in PMRA; (12) average/$1\sigma$ spread in PMDEC; (13) average/$1\sigma$ spread in heliocentric radial velocity; (14) number of available radial velocities from Gaia DR2, APOGEE-2, and GALAH DR3.}

\end{deluxetable*}

\subsection{Radial Velocities}

    To study the 3D kinematics of both OBP-near and Briceño-1, we require radial velocity (RV) measurements. We first cross-match the stars from OBP-near and Briceño-1 with the Sloan Digital Sky Survey IV \citep[SDSS-IV;][]{majewski2017} APOGEE-2 \citep{blanton2017, cottle2018} adopting \verb|SYNTHVHELIO_AVG|\footnote{An average of the RV measurements from cross-correlating to the best-fit synthetic spectrum.} as the RV measurement and \verb|SYNTHVERR| as their corresponding errors. Additionally, we include more RVs by cross-matching the two groups to GALAH DR3 \citep{bunder2020}. When a star has RV measurements in multiple surveys, we adopt the RV with the highest S/N, leading to a preference for APOGEE-2 RVs, followed by GALAH DR3, and then \textit{Gaia} DR2. There is no apparent systematic shift between APOGEE-2 and \textit{Gaia} DR2 RVs, however there is a positive shift in GALAH DR3 RVs by $\sim 0.7$ km~s$^{-1}$. We subtract this value from GALAH DR3 RVs to be inline with the \textit{Gaia} and APOGEE-2 RVs. 
    
    There is a large spread in the RV distribution, likely due to spectroscopic binaries in the sample. A CMD inspection of the two groups' stars with discrepant RVs reveals that many are O and B-type stars. According to \cite{chini2012}, the binary rate for stars with mass $>16~M_{\odot}$ is $\sim 80\%$ and drops to $\sim 20\%$ for stars with mass $\sim 3~M_{\odot}$. We exclude these stars by restricting the RVs to the range  17 $<$ RV $<$ 26 km s$^{-1}$. Overall, a total of 221 stars with accurate RVs remain in the sample of both groups.

\subsection{Positions and Kinematics}
    Knowing the positions and velocities of the two groups, we can constrain their dynamical evolution. We convert parallax to distance using $d = 1/\varpi$. We adopt the Astropy python package \citep{astropy2018} and convert the sky coordinates, distances, proper motions, and radial velocities of both groups' stars to 3D Galactic cartesian coordinates ($X, Y, Z$) pc and Heliocentric velocity components to ($U, V, W$) km~s$^{-1}$. A correction is made to the solar motion by subtracting  $(U, V, W)_{\odot} = (11.1, 12.24, 7.25)$ km~s$^{-1}$ from the stars' velocities \citep{schonrich2010} to obtain $(u,v,w)$ velocities with respect to the LSR. We also subtract Galactic rotation values at each star's Galactocentric radius using the \verb|MilkyWay2014| potential from the \verb|Galpy| python package \citep{bovy2015}. The difference in galactic rotation across the two groups is modest ($\sim 0.006$ km~s$^{-1}$~pc$^{-1}$). A correction to the proper motions' perspective contraction is made using Equation 13 from \cite{vanLeeuwen2008}.

    Errors on the positions and velocities are estimated using a Monte Carlo approach. The astrometric and RV errors are assumed to be normally distributed and uncorrelated given the proximity of Orion and the region's data quality. For a given star, Gaussian distributions are created for each measurement with the input errors corresponding to the distributions' standard deviations and the observed values corresponding to the mean. These distributions are sampled in parallel 10,000 times and transformed from spherical to Heliocentric Galactic cartesian coordinates. The standard deviations of each dimension's resulting distribution are then calculated and stored as their corresponding errors.  

    Finally, we define our study's reference frame by subtracting the median $(u,v,w)$ motions from the combined groups and adopting the notation $(X, Y, Z, v_x, v_y, v_z)$.

\subsection{Age Estimates}
    \label{sec:ages}
    To estimate the ages of Brice{\~n}o-1 and OBP-near, we fit isochrones to the groups' CMDs using their \textit{Gaia} EDR3 photometry. We use the latest PARSEC\footnote{\url{http://stev.oapd.inaf.it/cgi-bin/cmd}} isochrones \citep{marigo2017} with the updated EDR3 passband definitions. Similar to previous work, we consider both a stellar population's age ($t$) and extinction ($A_V$) as free parameters when fitting the models with an age range of $1 < t < 40 \text{ Myr}$ and a step size of 0.05 Myr and an extinction range of $0.0 < A_V < 1.0 \,\text{mag}$ with a step size of $0.1 \text{ mag}$ \citep{zari2019}. For simplicity, we assume a solar metal fraction of $Z = 0.0158$.
    
    The isochrone fitting method follows a standard maximum log-likelihood estimation procedure. Assuming the errors are Gaussian, we measure the likelihood of a star with a given mass, $m$, to come from an isochrone with parameters, $\boldsymbol{\theta} = (t, A_V)$ as: 
    
    \begin{equation}
        \label{log likelihood}
        \text{ln}\left(L(\boldsymbol{\theta},m)\right) = \sum_{i=1}^{n} \text{ln}\left(\frac{1}{(2\pi)^{1/2}\sigma_{i}}\right) - \frac{\chi^2}{2}
    \end{equation}
    
    \noindent where:
    
    \begin{equation}
        \label{chi2}
        \chi^2 = \sum_{i=1}^{n}\left(\frac{M_{g_i}^{\text{obs}} - M_{g_i}(\boldsymbol{\theta},m)}{\sigma_i}\right)^2
    \end{equation}
    
    \indent A quality cut is applied to the \textit{Gaia} photometry following \verb|phot_g_mean_flux_over_error| $> 20$ and \verb|phot_rp_mean_flux_over_error| $> 20$. We interpolate the isochrones to match the stars' \textit{Gaia} $G-G_{RP}$ color. $M_{g_i}$ is the absolute magnitude calculated from the \textit{G} band photometry and parallax. $M_{g_i}(t,m)$ is the absolute magnitude of the interpolated isochrone point corresponding to the same color as the observed star. The best-fit age (hereafter $t_{iso}$) is found by maximizing Eq. \ref{log likelihood} (minimizing Eq. \ref{chi2}) where the summation is performed across all CMD points of the group. Age uncertainties are adopted by the interquartile range of the likelihood distribution marginalized over ${{A}_{V}}$. These statistical uncertainties do not consider the biases that contribute to isochrone fitting: unresolved binaries, model uncertainties, and differential extinction. An unresolved binary sequence is apparent in both groups and is likely biasing the ages towards younger values. 

\begin{deluxetable*}{ccccccccccccc}[!hbt]
  \tablewidth{\columnwidth}
  \tabletypesize{\footnotesize}
    \tablecaption{SNN Catalog\label{tab:snn_catalog}}
    
    \tablehead{\colhead{Label} & \colhead{ID} & \colhead{$\alpha$} & \colhead{$\delta$} & \colhead{$\varpi$} & \colhead{$\mu_{\alpha}$} & \colhead{$\mu_{\delta}$} &
    \colhead{RV$_{\text{DR2}}$} & \colhead{RV$_{\text{A-2}}$} & 
    \colhead{RV$_{\text{GALAH}}$} &
    \colhead{p}
    \\
    & & (deg) & (deg) & (mas) & (mas~yr$^{-1}$) & (mas~yr$^{-1}$) & (km~s$^{-1}$) & (km~s$^{-1}$) & (km~s$^{-1}$) & ($\%$)
    \\
    (1) & (2) & (3) & (4) & (5) & (6) & (7) & (8) & (9) & (10) & (11)}
    
    \startdata
    1 & 3234338934867990528 & $81.4\pm0.1$ & $2.4\pm0.1$ & $2.8\pm0.2$ & $1.2\pm0.2$ & $0.2\pm0.1$ & – & – & – & 0.07 \\
    1 & 3234341030811875840 & $81.0\pm0.1$ & $2.3\pm0.1$ & $2.7\pm0.1$ & $1.2\pm0.2$ & $0.2\pm0.1$ & – & – & – & 0.06 \\
    1 & 3234341546207980032 & $81.1\pm0.1$ & $2.3\pm0.1$ & $2.8\pm0.1$ & $1.5\pm0.1$ & $-0.8\pm0.1$ & – & – & – & 0.07 \\
    1 & 3221860994017392128 & $81.2\pm0.1$ & $0.7\pm0.1$ & $2.8\pm0.1$ & $1.5\pm0.1$ & $-0.1\pm0.1$ & – & – & – & 0.18 \\
    1 & 3221860439965445120 & $81.2\pm0.2$ & $0.6\pm0.2$ & $3.0\pm0.2$ & $1.6\pm0.3$ & $0.2\pm0.2$ & – & – & – & 0.04 \\
    \enddata

    \tablecomments{Catalog of stars recovered as group members by SNN in this work with only the first five rows displayed. Data will be available for download through the online version of the published article. In column (1) we identify a star's SNN label corresponding to Table \ref{tab:summary_table}; (2) the \textit{Gaia} EDR3 source ID; (3) RA; (4) DEC; (5) parallax; (6) proper motion in RA; (7) proper motion in DEC; (8) \textit{Gaia} DR2 radial velocity; (9) APOGEE-2 radial velocity; (10) GALAH DR3 radial velocity; (11) SNN probability.}
    
\end{deluxetable*}

\section{Results}
\label{sec:results}

\begin{figure*}[!hbt]
    \centering
    \makebox[\textwidth]{\includegraphics[width=.75\paperwidth]{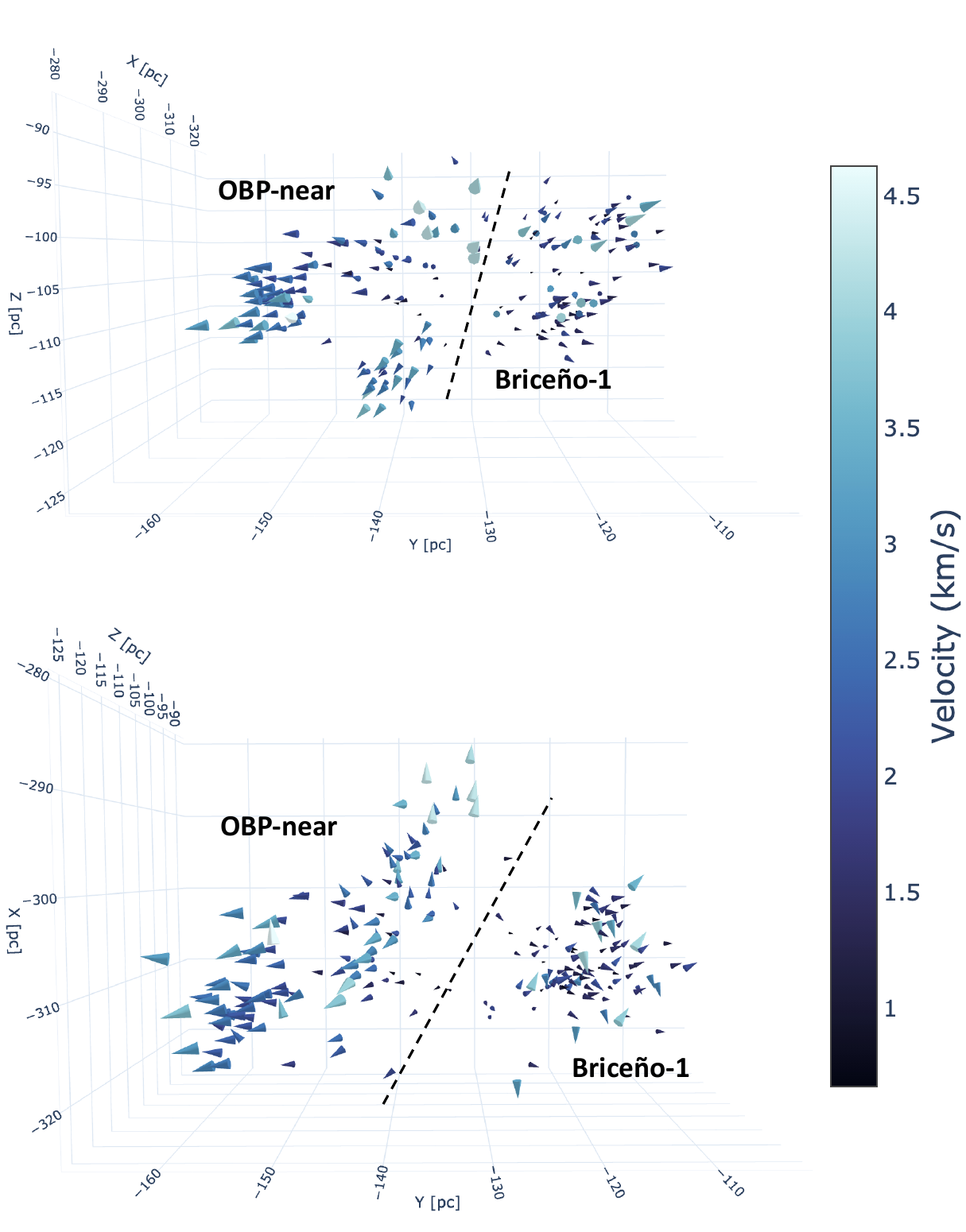}}
    \caption{3D kinematics of stars with available RVs in OBP-near and Briceño-1 in Heliocentric Galactic cartesian coordinates after subtracting the combined average of their kinematics in each dimension. Larger, lighter-colored cones represent faster-moving stars, and smaller, darker-colored cones represent slower-moving stars. The directions of the cones' apexes represent the directions of the stars' motions. The dashed line in each panel marks the spatial division of the two groups from that viewing angle. An interactive version is available \href{https://faun.rc.fas.harvard.edu/czucker/Paper\_Figures/expanding\_orion.html}{here}, or in the online version of the published article.}
    
    \label{fig:vectors}
\end{figure*}

\begin{figure}[hbt!]
        \centering
        \includegraphics[width = \linewidth]{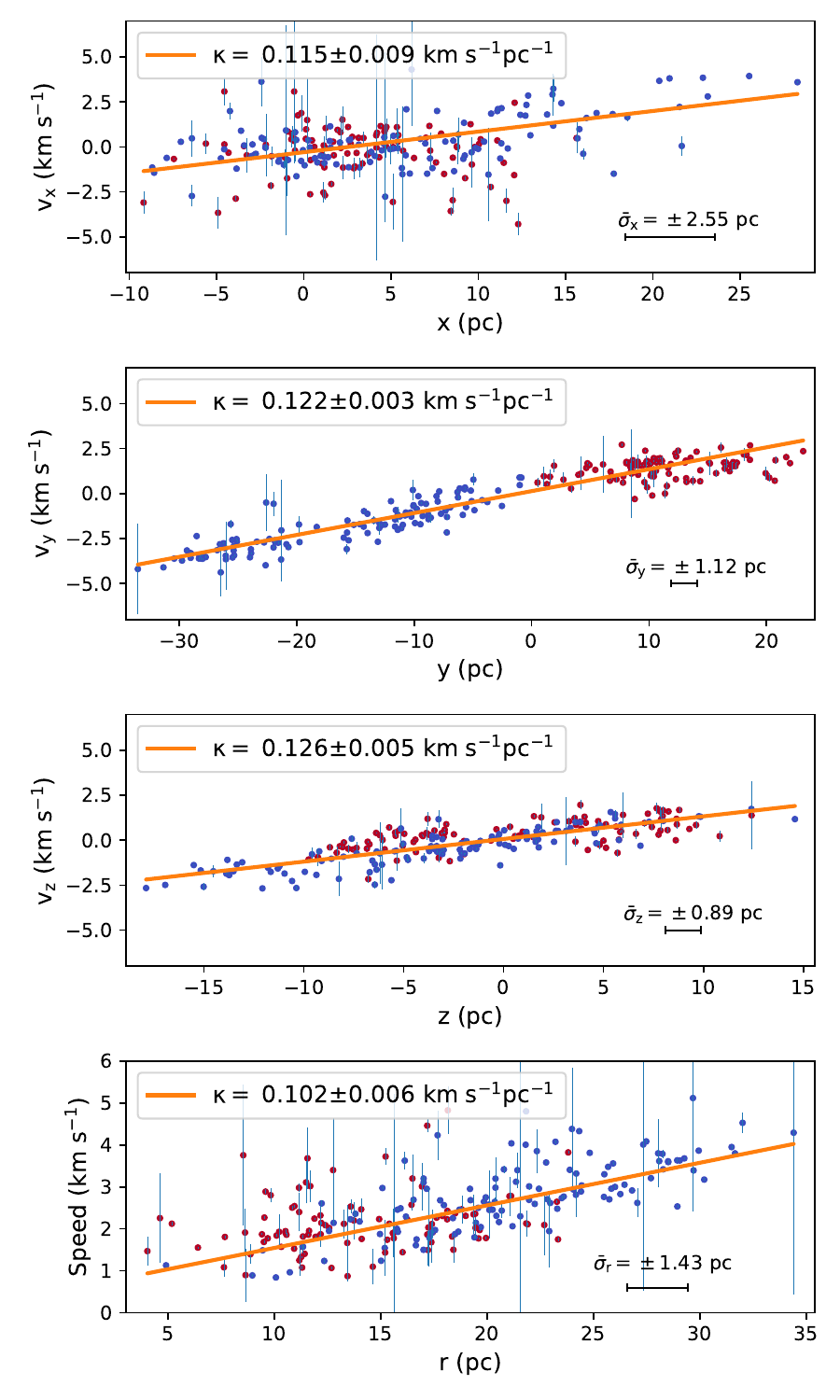}
        \caption{Position-velocity profiles of stars in the combined reference frame of OBP-near (blue points) and Briceño-1 (red points). Each panel shows the  position-velocity profile for a given Cartesian component. Regression lines (orange) are fit to the combined profiles of OBP-near and Briceño-1 in order to estimate the expansion slope, $\kappa$, which is quoted in the legends.  Individual velocity errors are displayed and the mean of the position errors is shown in the lower right-hand corner of each panel.} 
        \label{fig:components}
    \end{figure}

    When viewed in three spatial dimensions, the positions and kinematics reveal that the two groups are expanding radially away from a common center (Figure \ref{fig:vectors}). 
    Figure \ref{fig:components} shows a correlation between the position and velocity for both stellar groups. The slopes of the linear fits to the position-velocity profiles are ($\kappa_{x}, \kappa_{y}, \kappa_{z}$) = ($0.115 \pm 0.009, 0.122 \pm 0.003, 0.126 \pm 0.005$) km~s$^{-1}$ pc$^{-1}$. Hence, stars further away from the groups' center are moving faster. A fit to the stars' speeds as a function of radius yields a slope of $\kappa_{r} = 0.102 \pm 0.006$ km~s$^{-1}$ pc$^{-1}$ where the center is estimated to be $(X,Y,Z) = (-312, -130,-105)$ pc. This center is found using a grid search which minimizes the residuals between the radius versus speed fit. Using the stars' average position would be naive since the they are inhomogeneously distributed in space. The median speed of the stars in both groups is $2.37$ km~s$^{-1}$ with a median error of $\pm 0.04$ km~s$^{-1}$ and a corresponding velocity dispersion of $\sim0.80$ km~s$^{-1}$. 
    
    These linear profiles show evidence for ballistic expansion, where the highest velocity stars have had sufficient time to travel further outwards from the center. A process where stars are accelerated outwards due to violent changes in the region's gravitational potential may also play a role in creating the observed expansion profile \citep{zamora2019}.

    The correlation is tighter in $y-v_{y}$ and $z-v_{z}$ than in $x-v_{x}$. A Kendall's $\tau$ correlation test yields $\tau_{x-v_x} = .20$, $\tau_{y-v_y} = .73$, and $\tau_{z-v_z} = .50$, each with a statistical significance of $p < 1 \times 10^{-5}$. We note that among the three directions, the x-axis is the most aligned with the line-of-sight direction. As illustrated in Figure \ref{fig:components}, this bias results in large uncertainties in the parallaxes and RV estimates for the correlation projected along this component.  
    
    The expansion time estimate can be inferred using the stars' velocities as a function of position. For each of the position-velocity slopes found in Figure \ref{fig:components} we calculate the expansion time, $t_{\text{exp}} = 1/\gamma\kappa$, where $\gamma = 1.0227$ pc~Myr$^{-1}$~km$^{-1}$~s and is used as a conversion factor to ensure the calculated timescale is in Myr. We find $t_{\text{exp,x}} = 8.58$ Myr, $t_{\text{exp,y}} = 8.00$ Myr, $t_{\text{exp,z}} = 7.72$ Myr, and $t_{\text{exp,r}} = 9.58$ Myr. When tracing the stars' positions back in time (assuming constant velocity), we find that the groups were most compact around $7.5$ Myr ago, with a distance of only 6 pc between the two groups' respective centers. 

\begin{figure}[hbt!]
        \centering
        \includegraphics[width = .4\textwidth]{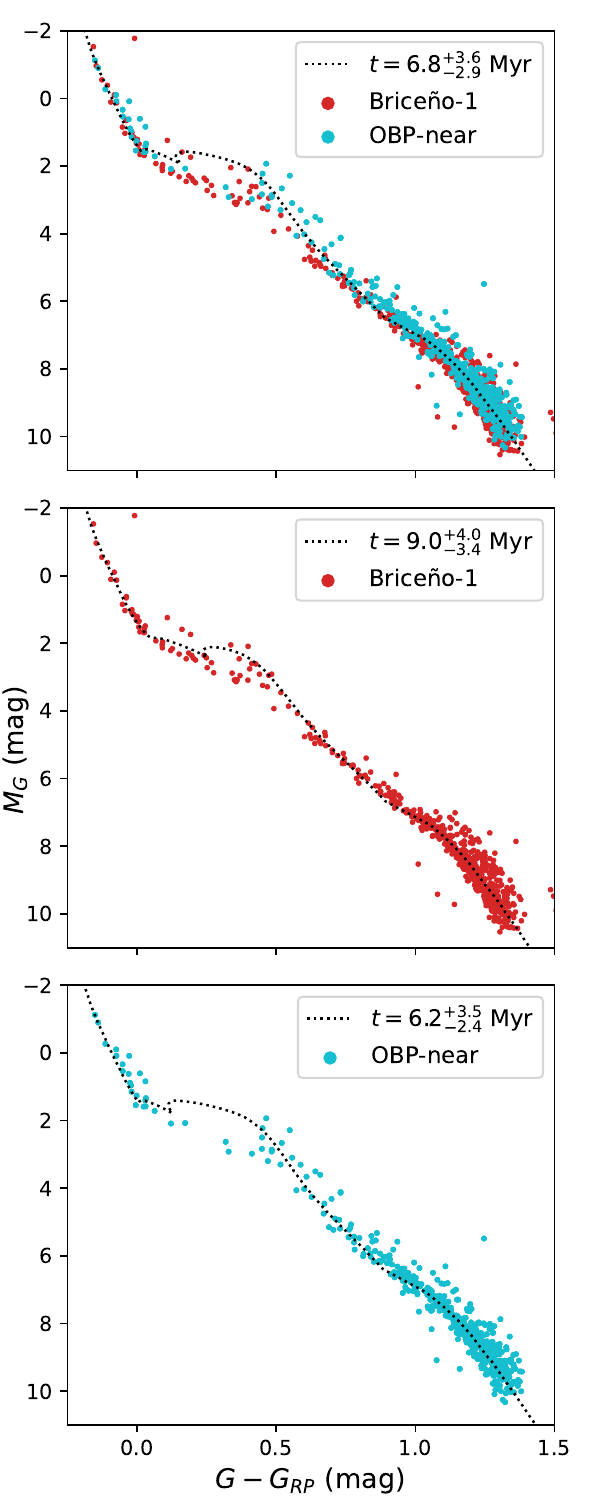}
        \caption{The CMDs of OBP-near (cyan dots) and Briceño-1 (red dots) stars: the top panel shows the combined CMD of both groups with their combined best-fit isochrone (dotted line); the middle panel shows the CMD of Briceño-1 and its best-fit isochrone; the bottom panel shows the CMD of OBP-near and its best-fit isochrone. The legend in each panel quotes the associated age of the isochrone fit and its uncertainty.}
        \label{fig:cmd}
    \end{figure}

    Figure \ref{fig:cmd} shows the color-magnitude diagram (CMD) of Brice{\~n}o-1 and OBP-near with their best-fit isochrone values, $t_{iso}$, and their associated uncertainties. From these isochrones, both groups are found to have an extinction of $A_{V} = 0.2 \text{ mag}$ with Briceño-1 having an age of $t_{iso} = 9.0^{+4.0}_{-3.4}$ Myr and OBP-near an age of $t_{iso} = 6.2^{+3.5}_{-2.4}$ Myr. The isochrone age when fitting to both groups combined is $t_{iso} = 6.8^{+3.6}_{-2.9}$ Myr. Given the large uncertainties, these values are in agreement with the $t_{\text{exp}}$ values. Briceño-1 appears to be $\sim$ 3 Myr older than OBP-near based on its $t_{iso}$ value and its offset on the CMD. However, the strong main-sequence overlap between the two groups makes this offset subtle. Note again that the $t_{iso}$ values are likely lower limits due to the bias from unresolved binaries in the sample. Additionally, Briceño-1 is positioned along the LOS, directly in front of the 20 Myr old group, ASCC20 \citep[][SNN7 in \citealp{chen2019}]{kos2019}, possibly biasing the isochrone fit for Briceño-1 towards an older value. However, SNN results do not indicate significant overlap from ASCC20. An investigation into the infrared colors of both Briceño-1 and OBP-near stars is presented in Appendix \ref{apx:ir-ccds} and points towards OBP-near indeed having a slightly younger age than Briceño-1, in agreement with their isochrone fits.

\section{Discussion}
\label{sec:discussion}

    \subsection{Feedback-induced Expansion}

\begin{figure*}[!hbt]
        \centering
        \includegraphics[width=\linewidth]{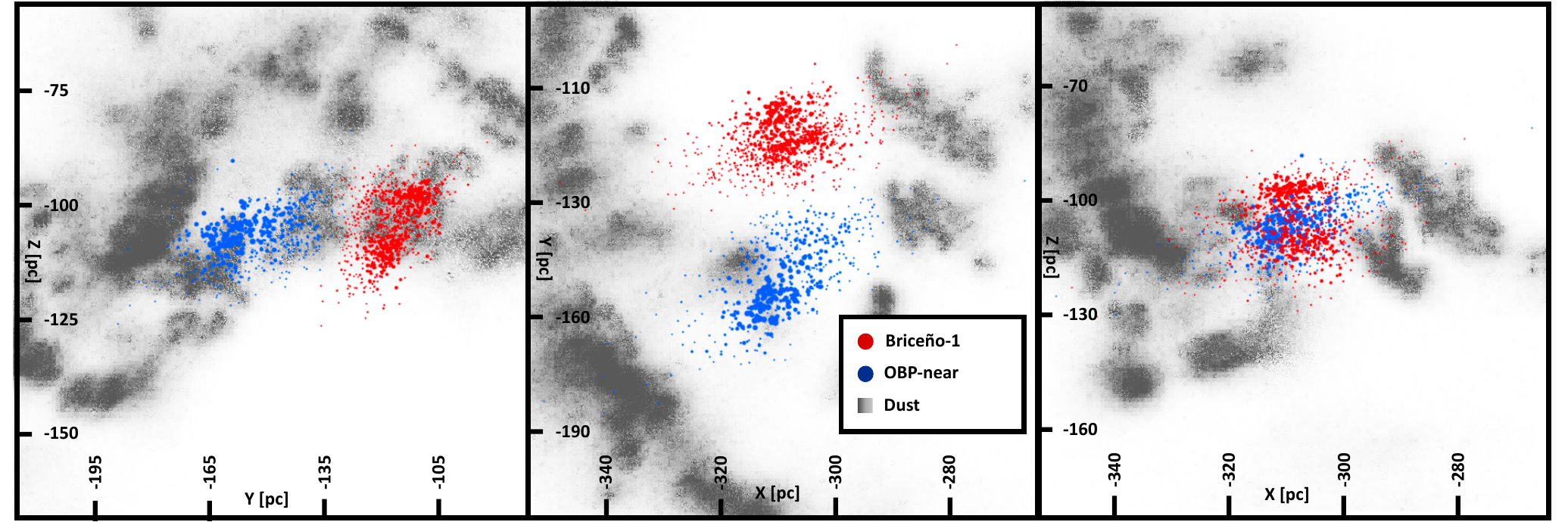}
        \caption{The spatial location of OBP-near and Briceño-1, overlaid on a 3D spatial map of interstellar dust \citep{leike2020}. As apparent in the top down X-Y view (middle panel), the OBP and Brice{\~n}o-1 groups lie at the center of a dust shell. The stars' sizes indicate their probability to be group members assigned by SNN. An interactive version of this figure is available \href{https://faun.rc.fas.harvard.edu/czucker/Paper_Figures/expanding\_orion\_3d\_dust.html}{here} or in the online version of the published article. The 3D dust visualization will fail to render in  Safari browser versions dating before MacOS Big Sur due to the lack of webp format support. In order to display dust, open the visualization in either the Chrome, Firefox, or Edge browser.}
        \label{fig:dust}
    \end{figure*}

    The radial expansion shown by OBP-near and Briceño-1 groups may be related to past SNe explosions or strong stellar winds which might have occurred at the core of Orion. Such winds and SNe would be expected to create cavities in the interstellar medium, manifesting as shells and bubbles \citep[see e.g.,][]{Smith_2019, Kim_2018}. Figure \ref{fig:dust} shows the spatial location of OBP-near and Briceño-1 (colored dots), as compared to a high-resolution 3D spatial map of the nearby interstellar medium derived from 3D dust mapping \citep{leike2020}. OBP-near and Briceño-1 lie at the center of a dust shell, which would indicate that past feedback events occurred within the two groups. Evidence of a dust shell is also seen in a complementary 3D dust map of the Orion A region, with the nearest part of the dust shell coincident with a previously undiscovered foreground cloud at a distance between $\approx 315-345$ pc \citep[see Figure 6 in][]{rezaei2020}. 
    Past studies have also found evidence for an expanding H~I shell that roughly surrounds the two groups' plane-of-sky positions but with a center more aligned with OBP-near \citep{chromey1989, ochsendorf2015}.
    
    One possible scenario to explain our observations suggests that stars which today identify as OBP-near and Briceño-1 formed approximately at the same time as one bound cluster at the core. A major feedback event dispersed the gas in the original molecular cloud after the stars formed. While the violent change of the gravitational potential might unbind the system and lead the stars to expand, it is unclear whether this mechanism can drive the stars into a ballistic expansion with the symmetry displayed in Figure \ref{fig:vectors}. As stated previously, rapid changes to the potential following gas/dust dispersal could aid in accelerating the stars \citep{zamora2019}. 
    
    Another theory posits that star associations may form in gas compressed layers of the expanding shell from shock fronts after SN events or other massive stellar feedback. In this scenario, the grouped stars will move at the same velocity as the expanding gas shell they formed in \citep{elmegreen1977}. Yet, it is unclear whether this process will lead to the expansion with the properties observed in Figures \ref{fig:vectors} and \ref{fig:components}. Such a scenario, where star formation is triggered sequentially, could explain an intrinsic age difference between OBP-near and Briceño-1; if feedback occurred closer to one side of the progenitor cloud, then the onset of star formation could have occurred at different times across different locations of the expanding gas shell. 

    \subsection{Comparison to Previous Works}

    A recent study performed a hierarchical clustering analysis on Orion and claimed that the entire complex shows clear radial expansion from a central region due to a SNe explosion \citep{kounkel2018, kounkel2020}. These studies recovered 190 individual groups, which were then recombined into larger groups as Orion A, B, C, D, and $\lambda$ Orion. Orion D was reported to consist of stars associated with the Orion OB1ab region and more diffuse populations just outside of it. While our analysis identifies stellar substructures in similar regions of Orion compared to \cite{kounkel2018}, it is clear that SNN applied to the improved EDR3 data yields different results: Of the Briceño-1 and OBP-stars identified in this work, a cross-match reveals that $43\%$ of these stars (679 stars) belong to Orion D of \cite{kounkel2018}. The groups ASCC20 \citep[SNN 7;][]{kos2019} and L1616 (SNN 8) recovered in this work partially cross-match to Orion D as well. Around $25\%$ of stars from OBP-d \citep[SNN 6;][]{kubiak2017}, OBP-b \citep[SNN 9;][]{kubiak2017}, and ome Ori \citep[SNN 14;][]{chen2019} overlap with Orion C.


    It was claimed that Orion D shows signs of expansion by analyzing proper motions with Gaia DR2 \citep{kounkel2018}; however, Orion D contains a mix of smaller groups that differ in age and kinematics. Isochrone fits to each of the SNN groups in this region indicate an age range between 6 and 22 Myr, with the oldest being ASCC20. These values suggest that Orion D has multiple stellar populations which presumably did not all form together.
    
    Furthermore, previous work placed all of these groups in a single, common reference frame and proposed that the entire complex is currently expanding from a central region due to a supernova event that occurred 6 Myr ago \citep{kounkel2020}. As part of the proposal, Orion C and D were once part of the same molecular cloud, split by the supernova explosion that separated them into two distinct regions. To test this hypothesis, we extend our analysis to all of the SNN groups recovered in this analysis utilizing their available radial velocities and estimating ages via the same method laid out in Section {\ref{sec:ages}}. 
    
\begin{figure*}[!ht]
        \centering
        \includegraphics[width = \linewidth]{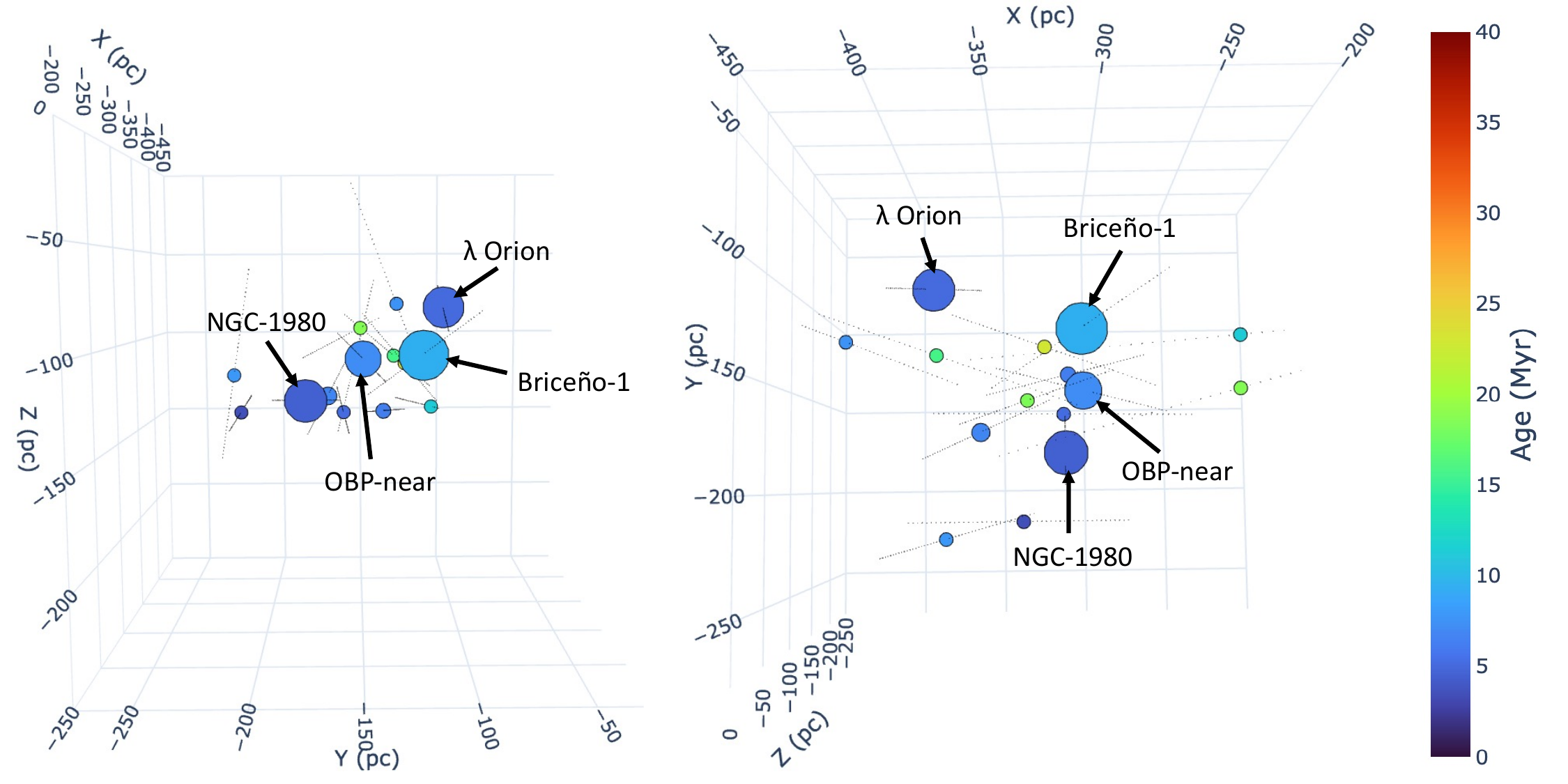}
        
        \caption{Interactive animation tracing the evolution of the SNN groups (Table \ref{tab:summary_table}) (colored dots) backwards ($-10$ Myr) and forwards ($+10$ Myr) in time in the Orion reference frame assuming a constant speed. Thin, dotted lines show the trace of each group's motion. Each panel in the static version shows a different view point of the groups' present-day positions. The groups' colors at $t=0$ represent their isochrone-derived ages and change correspondingly to reflect their ages at different time steps. Sizes indicate the number of stars they contain. Throughout the animation, filled dots become open dots if the chosen time is prior to a group's formation time.  The four most massive groups are indicated by the arrows and labels. Moving the time slider left moves the groups back in time and to the right moves them forwards in time. The interactive animation can be found \href{https://faun.rc.fas.harvard.edu/czucker/Paper\_Figures/orion\_movie.html}{here}, or in the online version of the published article.} 
        \label{fig:SNN_movie}
    \end{figure*}

    We follow the SNN groups' motion in a common reference frame backwards/forwards in time $\pm 10$ Myr (see Figure \ref{fig:SNN_movie}) and find that OBP-near and Briceño-1 are the only groups that clearly trace back to a common center. The motion of L1616 (N $=158$ stars) indicates that it might be part of the expansion. \cite{josefa2020} analyzes the complex's gas dynamics and finds L1616 to be part of an expanding region, however the stars they study are younger and might not match the stars of L1616 recovered by SNN. While the velocities of L1616's stars from SNN are a factor of $\sim 3$ too small to be consistent with the expansion profile of OBP-near and Briceño-1 (Figure \ref{fig:components}), this may be due to L1616 having a greater distance from the source of feedback during its onset. NGC 1980 partially overlaps with the Orion A region of \cite{kounkel2018}. However, our analysis shows that it does not originate from the expansion center when followed back in time. Its direction of motion does elicit the possibility that it has formed from the onset of triggered star formation due to feedback as originally proposed in \cite{kounkel2020} and \cite{josefa2020}. Note that there is an open debate questioning whether NGC 1980 is a separate, foreground population to the ONC group reported in \citet{kounkel2018} (see \citealp{alves2012} and \citealp{fang2017} for further discussion).
    
    Orion X \citep{bouy2015} was recovered as a low-stability group by SNN in \cite{chen2019}, however it is not recovered in this work. It is proposed as the origin of gas expansion in \cite{josefa2020} and is proximal to Briceño-1 and OBP-near (40-100 pc), further suggesting a connection between their results and the expansion observed here. However, the age of the gas expansion may be younger than the expansion of this work's stars, pointing towards a more complicated feedback scenario. Future investigations of the connection between Orion's gas and stellar kinematics might point to more precise past locations of stellar feedback in Orion.
    
    The SNN analysis reveals that groups with similar ages as Orion C and the $\sim 20$ Myr old groups overlapping with Orion D originate away from Orion's present-day center. Hence, signs of expansion are most evident in OBP-near and Brice{\~n}o-1 while the other groups show a more complex dynamical history. The identification of $> 15$ Myr old stars in Orion by \cite{kos2019} and \cite{jerabkova2019} and the 6D phase-space analysis here (Figure \ref{fig:SNN_movie}), elicits the possibility that leftover gas from an older epoch of star formation influenced the assembly of younger populations in the complex.

\section{Conclusions}
\label{sec:conclusion}

The astrometric measurements of the \textit{Gaia} mission provided in DR2 and EDR3, supplemented by APOGEE-2 and GALAH DR3, have enabled a systematic study of the 3D kinematics of OBP-near and Brice{\~n}o-1, two stellar groups at the core of the Orion complex. This work shows direct evidence of ballistic expansion occurring locally at the core of this region. 

While previous work found some evidence of coherent motions in Orion, this study shows compelling evidence of stellar expansion in a massive star-forming region. The stars are currently located at the center of a dust shell, suggesting that their expansion links to stellar feedback events. However, the process driving the relatively symmetric radial expansion shown in this study remains unclear. Upcoming work (Foley et al. in prep) will present further evidence for the connection between this radial expansion and stellar feedback. Future {\it Gaia} data releases are anticipated to significantly improve the astrometry of stars, which will allow for more accurate kinematic analyses of the entire Orion complex. Forthcoming observations and numerical studies of the interstellar medium will also help shed light on the relationship between feedback events and stellar dynamics

\section{Acknowledgments}

The authors are grateful to B. Elmegreen and J. Gallagher for insightful suggestions. This work has used data from the European Space Agency (ESA) mission Gaia (\url{https://www.cosmos.esa.int/gaia}), processed by the Gaia Data Processing and Analysis Consortium (DPAC; \url{https://www.cosmos.esa.int/web/gaia/dpac/consortium}).Funding for the DPAC has been provided by national institutions, in particular the institutions participating in the Gaia Multilateral Agreement. This project was developed in part at the 2019 Santa Barbara Gaia Sprint, hosted by the Kavli Institute for Theoretical Physics at the University of California, Santa Barbara, and by the  2020 EDR3 Unboxing Gaia Sprint, hosted by the Center for Computational Astrophysics of the Flatiron Institute in New York City.

Funding for the Sloan Digital Sky Survey IV has been provided by the Alfred P. Sloan Foundation, the U.S. Department of Energy Office of Science, and the Participating Institutions. SDSS-IV acknowledges support and resources from the Center for High Performance Computing  at the University of Utah. The SDSS website is www.sdss.org.

SDSS-IV is managed by the Astrophysical Research Consortium for the Participating Institutions of the SDSS Collaboration including the Brazilian Participation Group, the Carnegie Institution for Science, Carnegie Mellon University, Center for Astrophysics | Harvard \& Smithsonian, the Chilean Participation Group, the French Participation Group, Instituto de Astrof\'isica de Canarias, The Johns Hopkins University, Kavli Institute for the Physics and Mathematics of the Universe (IPMU) / University of Tokyo, the Korean Participation Group, Lawrence Berkeley National Laboratory, Leibniz Institut f\"ur Astrophysik Potsdam (AIP), Max-Planck-Institut f\"ur Astronomie (MPIA Heidelberg), Max-Planck-Institut f\"ur Astrophysik (MPA Garching), Max-Planck-Institut f\"ur Extraterrestrische Physik (MPE), National Astronomical Observatories of China, New Mexico State University, New York University, University of Notre Dame, Observat\'ario Nacional / MCTI, The Ohio State University, Pennsylvania State University, Shanghai Astronomical Observatory, United Kingdom Participation Group, Universidad Nacional Aut\'onoma de M\'exico, University of Arizona, University of Colorado Boulder, University of Oxford, University of Portsmouth, University of Utah, University of Virginia, University of Washington, University of Wisconsin, Vanderbilt University, and Yale University.

\textit{Software}: Astropy \citep{astropy2018}, NumPy \citep{van2011}, Plotly, Matplotlib \citep{hunter2007}, Glue \citep{robitaille2017}, and Galpy \citep{bovy2015}.

\begin{appendix}
\section{Infrared colors of OBP-near and Briceño-1} \label{apx:ir-ccds}

    \begin{figure*}[!ht]
        \centering
        \includegraphics[width = \linewidth]{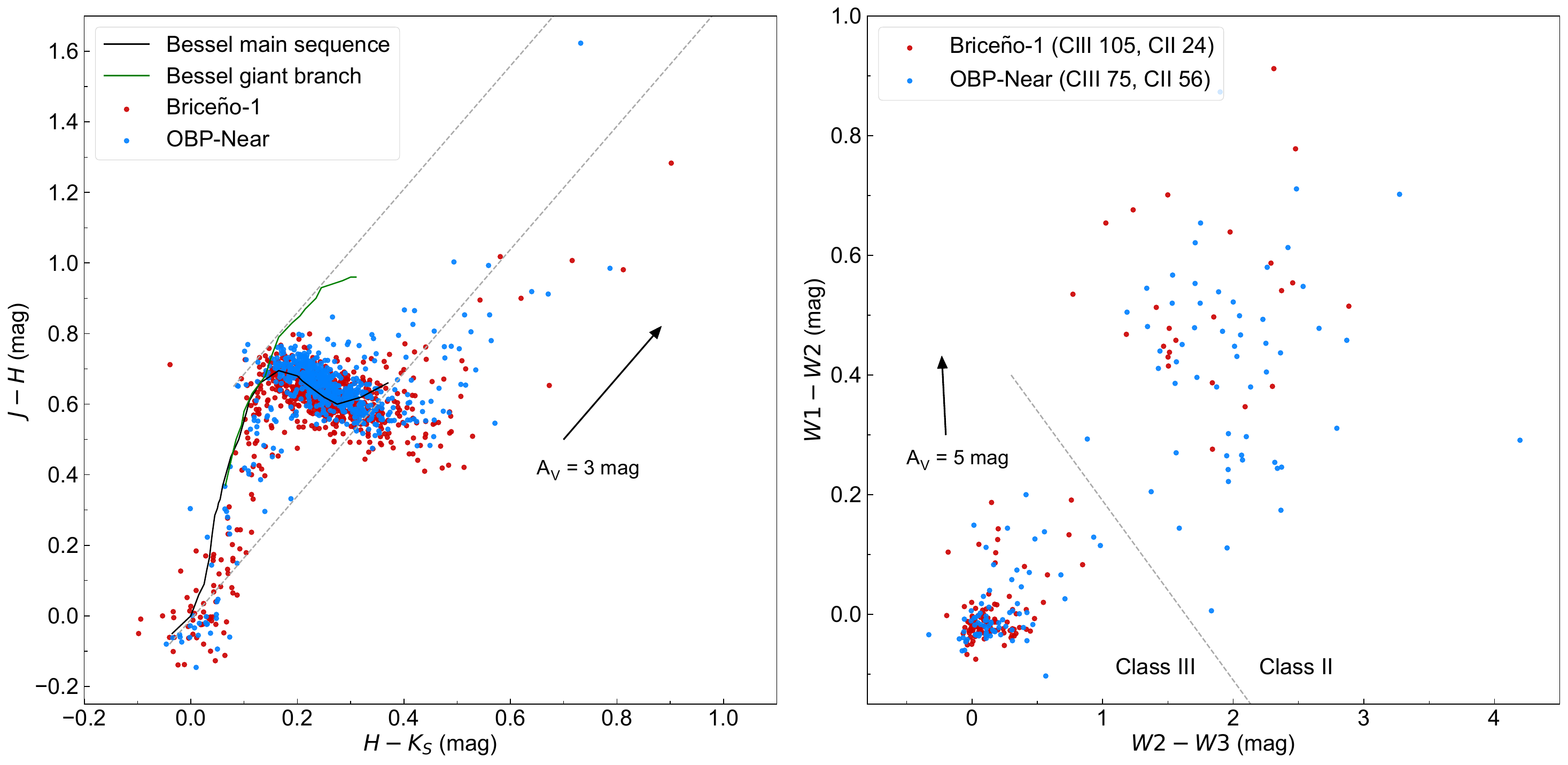}
        \caption{Infrared color-color diagrams for stellar members in Brice{\~n}o-1 (red) and OBP-near (blue).
        \textit{Left:} 2MASS near-infrared JHKs color-color diagram. Overplotted are the main sequence (black curve) and giant branch (green curve) from \cite{bessell1988} for reference. An extinction vector of length $A_V = 3$\,mag is shown as a black arrow, indicating the direction of reddening for extincted sources. The two gray dashed lines show the reddening band, plotted parallel to the extinction vector and above the main sequence.
        \textit{Right:} WISE mid-infrared color-color diagram. An extinction vector of length $A_V = 5$\,mag is shown as black arrow. The gray dashed line separates the groups roughly into Class II and Class III YSOs. The number of sources that fall in each class per group is given in the legend.}
        \label{fig:ir-ccds}
    \end{figure*}

Figure \ref{fig:ir-ccds} shows two infrared color-color diagrams to investigate how extinction is influencing the photometry of the two groups by using near-infrared photometry, and to investigate their evolutionary status by using mid-infrared photometry. For this analysis, we use near-infrared photometry, J ($1.25~\mu\text{m}$), H ($1.65~\mu\text{m}$), and K$_\mathrm{S}$ ($2.16~\mu\text{m}$), from the Two Micron All-Sky Survey \citep[2MASS;][]{skrutskie2006}, and mid-infrared photometry, W1 ($3.4~\mu\text{m}$), W2 ($4.6~\mu\text{m}$), and W3 ($12~\mu\text{m}$), from the Wide-field Infrared Survey Explorer \cite[WISE, AllWISE catalog;][]{cutri2013}, retrieved form \url{https://irsa.ipac.caltech.edu}. To combine the 2MASS and WISE catalogs with the Gaia EDR3 astrometry, we used a 1'' cross-match radius, only keeping the closest nearest neighbor. 

The left panel in Fig.~\ref{fig:ir-ccds} shows the 2MASS near-infrared color-color diagram, highlighting which sources in the two groups show significant signs of reddening, caused either by foreground extinction or by a circumstellar disk. In particular, sources above the main sequence, located within the reddening band, are thought to be dominated by foreground extinction. Only sources that pass the following quality criteria were used for the diagram:

\begin{equation} \label{equ:2mass}
\begin{aligned}
  &  \mathrm{j\_sig}, \, \mathrm{h\_sig}, \, \mathrm{k\_sig} < 0.1 \text{ mag},  \\
  &  \mathrm{j\_snr}, \, \mathrm{h\_snr}, \, \mathrm{k\_snr} > 10.  \\
\end{aligned}
\end{equation}

With these cuts there are 850 and 642 sources left in Brice{\~n}o-1 and OBP-near, respectively. The majority of the sources in the two groups (more than 95\%) show no or very little reddening, with no significant difference between the groups. This is expected, since the groups lie in-front of the major cloud complexes in Orion and it would correlate with the low average extinction found  during isochrone fitting. There are a few sources showing extinctions as high as about $A_V=3$ to 8\,mag, while most of these are classified as Class II YSOs based on their mid-infrared color (see below), hence the reddening could be caused by circumstellar dust. The source showing the highest reddening in Fig.~\ref{fig:ir-ccds} (2MASS J05414177-0151456 in OBP-near) is projected on top of the young embedded cluster NGC\,2024 in Orion\,B and might rather belong to this cluster, since some contamination by other groups in our SNN selected samples can not be ruled out.

The right panel in Fig.~\ref{fig:ir-ccds} shows a WISE mid-infrared color-color diagram. To use near- to mid-infrared photometry to distinguish evolutionary classes within a young stellar population based on the infrared excess determined from the spectral energy distribution (SED) is a proofed tool since the 80s \cite[e.g.,][]{lada1987, lada2006, evans2009, josefa2019}. This allows one to estimate which of the young stellar objects (YSOs) are still embedded in an envelope (Class\,I), are pre-main-sequence stars surrounded by a circumstellar disk (Class\,II), or which are pre-main-sequence stars that have already dissipated their envelope and disk and are on their way to the main-sequence (Class\,III). The latter class can not be strictly separated from main-sequence stars using solely mid-infrared colors, however, since we are dealing in this study with populations younger than about 10\,Myr, the classification into Class\,III for sources with no clear infrared excess is a feasible approach. We like to note that the most massive stars of the two populations have likely already reached the main-sequence while still being young. Regarding Class\,I protostars, there are likely no protostars included in the samples which show IR-excess since they tend to show redder colors beyond the borders of the displayed axes, hence sources with IR-excess will be labeled as Class\,II. To get reliable WISE photometry, several quality criteria have been applied to avoid erroneous photometry and contamination by extended emission, to which especially the W3 passband is susceptible to. The quality criteria are as follows:

\begin{equation} \label{equ:wise}
\begin{aligned}
  &  \mathrm{w1snr, w2snr, w3snr} > 10,  \\
  &  \mathrm{w1rchi2, w2rchi2, w3rchi3} < 20,  \\
  &  \mathrm{w3mag\_1-w3mag\_6} < 2\,\mathrm{mag}. \\
\end{aligned}
\end{equation}

For more details on the WISE quality criteria see \cite{josefa2019}.
After applying these quality criteria, the displayed sources in Figure 7 only represent a rather small fraction (about 17$\%$) of the SNN selected members. This is due to the inferior resolution and sensitivity of WISE compared to Gaia or 2MASS photometry. In particular, with these cuts there are 129 and 131 sources left in Brice{\~n}o-1 and OBP-near, respectively. The Class\,II YSOs are separated from Class\,III and main-sequence stars with
\begin{equation} \label{equ:class-cut}
    \mathrm{W1-W2} > -0.3 \cdot (W2-W3-0.3)+0.4.
\end{equation}
This separation is similar to class definitions in the literature based on WISE colors as discussed, for example, in \cite{koenig2014}.
With this we get a Class\,II disk fraction of about 19\% and 43\% for Brice{\~n}o-1 and OBP-near, respectively (only using the sources shown in Fig.~\ref{fig:ir-ccds} in the right panel). This difference in disk fraction suggests that OBP-near is at a younger evolutionary status compared to Brice{\~n}o-1, in agreement with the difference in ages as obtained via isochrones in Sect.~\ref{sec:results}.

\end{appendix}


\begin{thebibliography}{}

\bibitem[Alves \& Bouy(2012)]{alves2012} Alves, J. \& Bouy, H.\ 2012, \aap, 547, A97. doi:10.1051/0004-6361/201220119

\bibitem[Alves et al.(2020)]{alves2020} Alves, J., Zucker, C., Goodman, A.~A., et al.\ 2020, \nat, 578, 237

\bibitem[Astropy Collaboration et al.(2018)]{astropy2018} Astropy Collaboration, Price-Whelan, A.~M., Sip{\H{o}}cz, B.~M., et al.\ 2018, \aj, 156, 123. doi:10.3847/1538-3881/aabc4f

\bibitem[Bally(2008)]{bally2008} Bally, J.\ 2008, Handbook of Star Forming Regions, Volume I, 459


\bibitem[Baumgardt \& Kroupa(2007)]{baumgardt2007} Baumgardt, H. \& Kroupa, P.\ 2007, \mnras, 380, 1589. doi:10.1111/j.1365-2966.2007.12209.x

\bibitem[Bessell \& Brett(1988)]{bessell1988} Bessell, M.~S. \& Brett, J.~M.\ 1988, \pasp, 100, 1134. doi:10.1086/132281

\bibitem[Blanton et al.(2017)]{blanton2017} Blanton, M.~R., Bershady, M.~A., Abolfathi, B., et al.\ 2017, \aj, 154, 28

\bibitem[Blaauw(1964)]{blaauw1964} Blaauw, A.\ 1964, \araa, 2, 213

\bibitem[Bouy \& Alves(2015)]{bouy2015} Bouy, H. \& Alves, J.\ 2015, \aap, 584, A26. doi:10.1051/0004-6361/201527058


\bibitem[Bovy(2015)]{bovy2015} Bovy, J.\ 2015, \apjs, 216, 29


\bibitem[Brice{\~n}o et al.(2007)]{briceno2007} Brice{\~n}o, C., Hartmann, L., Hern{\'a}ndez, J., et al.\ 2007, \apj, 661, 1119

\bibitem[Brown et al.(1994)]{brown1994} Brown, A.~G.~A., de Geus, E.~J., \& de Zeeuw, P.~T.\ 1994, \aap, 289, 101


\bibitem[Buder et al.(2020)]{bunder2020} Buder, S., Sharma, S., Kos, J., et al.\ 2020, arXiv:2011.02505

\bibitem[Cantat-Gaudin et al.(2019)]{cantat2019} Cantat-Gaudin, T., Jordi, C., Wright, N.~J., et al.\ 2019, \aap, 626, A17. doi:10.1051/0004-6361/201834957

\bibitem[Chen et al.(2018)]{chen2018} Chen, B., D'Onghia, E., Pardy, S.~A., et al.\ 2018, \apj, 860, 70. doi:10.3847/1538-4357/aac325

\bibitem[Chen et al.(2020)]{chen2019} Chen, B., D'Onghia, E., Alves, J., et al.\ 2020, \aap, 643, A114. doi:10.1051/0004-6361/201935955

\bibitem[Chini et al.(2012)]{chini2012} Chini, R., Hoffmeister, V.~H., Nasseri, A., et al.\ 2012, \mnras, 424, 1925. doi:10.1111/j.1365-2966.2012.21317.x

\bibitem[Chromey et al.(1989)]{chromey1989} Chromey, F.~R., Elmegreen, B.~G., \& Elmegreen, D.~M.\ 1989, \aj, 98, 2203. doi:10.1086/115289

\bibitem[Cottle et al.(2018)]{cottle2018} Cottle, J. 'Neil ., Covey, K.~R., Su{\'a}rez, G., et al.\ 2018, \apjs, 236, 27. doi:10.3847/1538-4365/aabada


\bibitem[Cropper et al.(2018)]{cropper2018} Cropper, M., Katz, D., Sartoretti, P., et al.\ 2018, \aap, 616, A5

\bibitem[Cutri et al.(2013)]{cutri2013} Cutri, R.~M., Wright, E.~L., Conrow, T., et al.\ 2013, Explanatory Supplement to the AllWISE Data Release Products, by R. M. Cutri et al.


\bibitem[Elmegreen \& Lada(1977)]{elmegreen1977} Elmegreen, B.~G. \& Lada, C.~J.\ 1977, \apj, 214, 725. doi:10.1086/155302

\bibitem[Evans et al.(2009)]{evans2009} Evans, N.~J., Dunham, M.~M., J{\o}rgensen, J.~K., et al.\ 2009, \apjs, 181, 321. doi:10.1088/0067-0049/181/2/321

\bibitem[Fang et al.(2017)]{fang2017} Fang, M., Kim, J.~S., Pascucci, I., et al.\ 2017, \aj, 153, 188. doi:10.3847/1538-3881/aa647b

\bibitem[Gaia Collaboration et al.(2016)]{gaia2016} Gaia Collaboration, Prusti, T., de Bruijne, J.~H.~J., et al.\ 2016, \aap, 595, A1. doi:10.1051/0004-6361/201629272

\bibitem[Gaia Collaboration et al.(2018)]{gaiaDR2} Gaia Collaboration, Brown, A.~G.~A., Vallenari, A., et al.\ 2018, \aap, 616, A1

\bibitem[Gaia Collaboration et al.(2020)]{gaiaEDR3} Gaia Collaboration, Brown, A.~G.~A., Vallenari, A., et al.\ 2020, arXiv:2012.01533

\bibitem[Geen et al.(2018)]{geen2018} Geen, S., Watson, S.~K., Rosdahl, J., et al.\ 2018, \mnras, 481, 2548. doi:10.1093/mnras/sty2439

\bibitem[Goodwin \& Bastian(2006)]{goodwin2006} Goodwin, S.~P. \& Bastian, N.\ 2006, \mnras, 373, 752. doi:10.1111/j.1365-2966.2006.11078.x

\bibitem[Gro{\ss}schedl et al.(2018)]{Grossschedl_2018} Gro{\ss}schedl, J.~E., Alves, J., Meingast, S., et al.\ 2018, \aap, 619, A106. doi:10.1051/0004-6361/201833901

\bibitem[Gro{\ss}schedl et al.(2019)]{josefa2019} Gro{\ss}schedl, J.~E., Alves, J., Teixeira, P.~S., et al.\ 2019, \aap, 622, A149. doi:10.1051/0004-6361/201832577

\bibitem[Gro{\ss}schedl et al.(2021)]{josefa2020} Gro{\ss}schedl, J.~E., Alves, J., Meingast, S., et al.\ 2021, \aap, 647, A91. doi:10.1051/0004-6361/202038913


\bibitem[Hillenbrand \& Hartmann(1998)]{hillenbrand1998} Hillenbrand, L.~A. \& Hartmann, L.~W.\ 1998, \apj, 492, 540. doi:10.1086/305076

\bibitem[Hills(1980)]{hills1980} Hills, J.~G.\ 1980, \apj, 235, 986. doi:10.1086/157703

\bibitem[Hunter(2007)]{hunter2007} Hunter, J.~D.\ 2007, Computing in Science and Engineering, 9, 90. doi:10.1109/MCSE.2007.55

\bibitem[Jeffries et al.(2011)]{jeffries_2011} Jeffries, R.~D., Littlefair, S.~P., Naylor, T., et al.\ 2011, \mnras, 418, 1948. doi:10.1111/j.1365-2966.2011.19613.x


\bibitem[Jerabkova et al.(2019)]{jerabkova2019} Jerabkova, T., Boffin, H.~M.~J., Beccari, G., et al.\ 2019, \mnras, 489, 4418. doi:10.1093/mnras/stz2315

\bibitem[Kim \& Ostriker(2018)]{Kim_2018} Kim, C.-G. \& Ostriker, E.~C.\ 2018, \apj, 853, 173. doi:10.3847/1538-4357/aaa5ff

\bibitem[Koenig \& Leisawitz(2014)]{koenig2014} Koenig, X.~P. \& Leisawitz, D.~T.\ 2014, \apj, 791, 131. doi:10.1088/0004-637X/791/2/131


\bibitem[Kos et al.(2019)]{kos2019} Kos, J., Bland-Hawthorn, J., Asplund, M., et al.\ 2019, \aap, 631, A166. doi:10.1051/0004-6361/201834710

\bibitem[Kounkel et al.(2018)]{kounkel2018} Kounkel, M., Covey, K., Su{\'a}rez, G., et al.\ 2018, \aj, 156, 84. doi:10.3847/1538-3881/aad1f1

\bibitem[Kounkel(2020)]{kounkel2020} Kounkel, M.\ 2020, \apj, 902, 122. doi:10.3847/1538-4357/abb6e8

\bibitem[Krause et al.(2020)]{krause2020} Krause, M.~G.~H., Offner, S.~S.~R., Charbonnel, C., et al.\ 2020, \ssr, 216, 64. doi:10.1007/s11214-020-00689-4

\bibitem[Kroupa et al.(2001)]{kroupa2001} Kroupa, P., Aarseth, S., \& Hurley, J.\ 2001, \mnras, 321, 699. doi:10.1046/j.1365-8711.2001.04050.x

\bibitem[Kroupa et al.(2018)]{kroupa2018} Kroupa, P., Je{\v{r}}{\'a}bkov{\'a}, T., Dinnbier, F., et al.\ 2018, \aap, 612, A74. doi:10.1051/0004-6361/201732151


\bibitem[Kubiak et al.(2017)]{kubiak2017} Kubiak, K., Alves, J., Bouy, H., et al.\ 2017, \aap, 598, A124

\bibitem[Kuhn et al.(2019)]{kuhn2019} Kuhn, M.~A., Hillenbrand, L.~A., Sills, A., et al.\ 2019, \apj, 870, 32. doi:10.3847/1538-4357/aaef8c

\bibitem[Kuhn et al.(2020)]{kuhn2020} Kuhn, M.~A., Hillenbrand, L.~A., Carpenter, J.~M., et al.\ 2020, \apj, 899, 128. doi:10.3847/1538-4357/aba19a

\bibitem[Lada(1987)]{lada1987} Lada, C.~J.\ 1987, Star Forming Regions, 115, 1

\bibitem[Lada \& Lada(2003)]{lada2003} Lada, C.~J. \& Lada, E.~A.\ 2003, \araa, 41, 57. doi:10.1146/annurev.astro.41.011802.094844

\bibitem[Lada et al.(2006)]{lada2006} Lada, C.~J., Muench, A.~A., Luhman, K.~L., et al.\ 2006, \aj, 131, 1574. doi:10.1086/499808

\bibitem[Lada et al.(2010)]{lada2010} Lada, C.~J., Lombardi, M., \& Alves, J.~F.\ 2010, \apj, 724, 687. doi:10.1088/0004-637X/724/1/687

\bibitem[Leike et al.(2020)]{leike2020} Leike, R.~H., Glatzle, M., \& En{\ss}lin, T.~A.\ 2020, \aap, 639, A138. doi:10.1051/0004-6361/202038169


\bibitem[Luri et al.(2018)]{luri2018} Luri, X., Brown, A.~G.~A., Sarro, L.~M., et al.\ 2018, \aap, 616, A9

\bibitem[Majewski et al.(2017)]{majewski2017} Majewski, S.~R., Schiavon, R.~P., Frinchaboy, P.~M., et al.\ 2017, \aj, 154, 94

\bibitem[Marigo et al.(2017)]{marigo2017} Marigo, P., Girardi, L., Bressan, A., et al.\ 2017, \apj, 835, 77

\bibitem[Ochsendorf et al.(2015)]{ochsendorf2015} Ochsendorf, B.~B., Brown, A.~G.~A., Bally, J., et al.\ 2015, \apj, 808, 111. doi:10.1088/0004-637X/808/2/111

\bibitem[Rezaei Kh. et al.(2020)]{rezaei2020} Rezaei Kh., S., Bailer-Jones, C.~A.~L., Soler, J.~D., et al.\ 2020, \aap, 643, A151. doi:10.1051/0004-6361/202038708

\bibitem[Robitaille et al.(2017)]{robitaille2017} Robitaille, T., Beaumont, C., Qian, P., et al.\ 2017, Zenodo


\bibitem[Rom{\'a}n-Z{\'u}{\~n}iga et al.(2019)]{roman2019} Rom{\'a}n-Z{\'u}{\~n}iga, C.~G., Roman-Lopes, A., Tapia, M., et al.\ 2019, \apjl, 871, L12. doi:10.3847/2041-8213/aafb06


\bibitem[Sch{\"o}nrich et al.(2010)]{schonrich2010} Sch{\"o}nrich, R., Binney, J., \& Dehnen, W.\ 2010, \mnras, 403, 1829

\bibitem[Sharma \& Johnston(2009)]{sharma2009} Sharma, S. \& Johnston, K.~V.\ 2009, \apj, 703, 1061. doi:10.1088/0004-637X/703/1/1061

\bibitem[Skrutskie et al.(2006)]{skrutskie2006} Skrutskie, M.~F., Cutri, R.~M., Stiening, R., et al.\ 2006, \aj, 131, 1163. doi:10.1086/498708

\bibitem[Smith et al.(2020)]{Smith_2019} Smith, R.~J., Tre{\ss}, R.~G., Sormani, M.~C., et al.\ 2020, \mnras, 492, 1594. doi:10.1093/mnras/stz3328

\bibitem[Tutukov(1978)]{tutukov1997} Tutukov, A.~V.\ 1978, \aap, 70, 57


\bibitem[Ward \& Kruijssen(2018)]{ward2018} Ward, J.~L. \& Kruijssen, J.~M.~D.\ 2018, \mnras, 475, 5659. doi:10.1093/mnras/sty117


\bibitem[Ward et al.(2020)]{ward2020} Ward, J.~L., Kruijssen, J.~M.~D., \& Rix, H.-W.\ 2020, \mnras, 495, 663. doi:10.1093/mnras/staa1056

\bibitem[Wright \& Mamajek(2018)]{wright2018} Wright, N.~J. \& Mamajek, E.~E.\ 2018, \mnras, 476, 381. doi:10.1093/mnras/sty207

\bibitem[Wright et al.(2019)]{wright2019} Wright, N.~J., Jeffries, R.~D., Jackson, R.~J., et al.\ 2019, \mnras, 486, 2477. doi:10.1093/mnras/stz870

\bibitem[van der Walt et al.(2011)]{van2011} van der Walt, S., Colbert, S.~C., \& Varoquaux, G.\ 2011, Computing in Science and Engineering, 13, 22. doi:10.1109/MCSE.2011.37


\bibitem[van Leeuwen(2009)]{vanLeeuwen2008} van Leeuwen, F.\ 2009, \aap, 497, 209. doi:10.1051/0004-6361/200811382


\bibitem[Zamora-Avil{\'e}s et al.(2019)]{zamora2019} Zamora-Avil{\'e}s, M., Ballesteros-Paredes, J., Hern{\'a}ndez, J., et al.\ 2019, \mnras, 488, 3406. doi:10.1093/mnras/stz1897


\bibitem[Zari et al.(2017)]{zari2017} Zari, E., Brown, A.~G.~A., de Bruijne, J., et al.\ 2017, \aap, 608, A148. doi:10.1051/0004-6361/201731309

\bibitem[Zari et al.(2019)]{zari2019} Zari, E., Brown, A.~G.~A., \& de Zeeuw, P.~T.\ 2019, \aap, 628, A123

\bibitem[Zucker et al.(2020)]{Zucker_2020} Zucker, C., Speagle, J.~S., Schlafly, E.~F., et al.\ 2020, \aap, 633, A51. doi:10.1051/0004-6361/201936145
 
\end{thebibliography}
\end{document}